\documentclass[twocolumn,prb,showpacs,preprintnumbers,amsmath,amssymb]{revtex4}

\usepackage{graphicx}
\usepackage{dcolumn}
\usepackage{bm}

\begin{document}



\title{Temperature dependent effective mass renormalization in 2D
  electron systems}

\author{S. Das Sarma}
\author{Victor M. Galitski}
\author{Ying Zhang}
\affiliation{Condensed Matter Theory Center, Department of Physics,
University of Maryland, College Park, MD 20742-4111}

\date{\today}

\begin{abstract}
We calculate, as a function of temperature and density, the
electron-electron interaction induced quasiparticle effective mass
renormalization in 2D electron systems within the leading-order
dynamically screened Coulomb interaction expansion. We find an
unexpected nonanalyticity and nonmonotonicity in the temperature
dependent effective mass with the renormalized mass linearly
increasing with temperature at low temperatures for all densities. 
\end{abstract}


\maketitle

\section{Introduction}
\label{sec:intro}

A key insight of the Landau Fermi Liquid theory is that interactions
in a Fermi system lead to the renormalization of the single particle
fermion mass giving rise to ``quasiparticles'' with renormalized
effective mass whose low energy behavior is qualitatively similar to
the corresponding noninteracting free particles~\cite{pines,AGD}. Then, 
various single particles properties, e.g. specific heat, density of
states, etc., in the interacting Fermi system are simply given, at
least in the leading-order theory, by replacing the bare (i.e. ``free
particle'') mass $m$ by the corresponding renormalized effective mass
$m^*$. In this paper we present the very {\it first}  microscopic
calculation of the {\it temperature dependent} effective mass
renormalization in an interacting 2D electron system (2DES), finding
in the process an unexpected nonanalytic and nonmonotonic behavior of
the effective mass $m^*(T)$ as a function of temperature in the
2DES. In particular, $m^*(T)$ first increases linearly with
temperature in a 2DES reaching a density dependent maximum around
$T/T_F \lesssim 0.1\, \mbox{---}\, 0.5$, where $T_F$ is the
noninteracting Fermi temperature, after which it decreases with
increasing temperature. This nonmonotonic behavior, in particular the
temperature induced enhancement of the 2DES quasiparticle effective
mass at low temperatures, is entirely unexpected because the naive
expectation is that quantum many-body electron-electron interaction
effects (underlying the effective mass renormalization phenomenon)
should decrease with increasing temperature since the high temperature
system is necessarily a classical system. The nonanalytic linear-$T$
dependence of $m^*(T)$ is also quite unexpected since the usual
fermionic Sommerfeld thermal expansion always results in a quadratic
temperature correction.  

Our work is partially motivated by the great deal of recent activity
in semiconductor-based 2DES, e.g. Si inversion layers, GaAs
heterostructures and quantum wells, etc. where the 2D carrier density
can be varied (by tuning an external gate voltage), modifying the
strength of the electron-electron interaction usually
measured~\cite{pines,AGD} by the dimensionless parameter $r_s = m e^2
/ ( \hbar^2 \sqrt{\pi n})$ with $n$ being the 2D carrier density and
$m$ the bare (i.e. band) mass. The $r_s$-parameter~\cite{pines} in 3D
metals (defined with respect to 3D densities) is typically $3-5$
whereas in semiconductor 2DES $r_s$ could vary from $1$ (or less) to
$20$ (or higher), depending on the specific semiconductor system and
carrier density being studied. Since the effective mass
renormalization scales with $r_s$ (small and large $r_s$ respectively
corresponding to weakly and strongly interacting electron systems),
one expects interesting and important many-body quasiparticle
renormalization in 2DES, particularly at large $r_s$. It is therefore
not surprising that the issue of the effective mass renormalization in
2DES has been extensively studied, both experimentally~\cite{smith}
and theoretically~\cite{ting,vinter,jalabert,schulze,ando} over the
last thirty years. All these theoretical studies of quasiparticle mass
renormalization have, however, been restricted to $T = 0$ both in the
2D~\cite{ting,vinter,jalabert,schulze,ando} and 3D~\cite{rice}
system. While this zero-temperature restriction makes perfect sense in
3D systems where the relevant Fermi temperature $T_F = E_F /k_B$
(defining the temperature scale for the electron system) is extremely
high ($T_F \sim 10^4 K$ in metals), it makes little sense for
extremely low density 2DES of current
interest~\cite{shashkin1,shashkin2,zhu} where $T_F \lesssim 1 K$,
making $T/T_F \sim 1$ in the experimental temperature range. The
temperature (and density) dependent effective mass renormalization
calculation presented in this paper therefore takes on additional
significance because a number of recent experiments have reported
large 2D effective mass renormalization~\cite{shashkin1,shashkin2,zhu}
at low densities and low temperatures. We note in this context that
the 2D effective mass renormalization $m^*/m$ in our finite 
temperature many-body theory is a function of two dimensionless
parameter $r_s$  ($\propto n^{-1/2}$) and $T/T_F$ ($\propto n^{-1}$,
since $k_B T_F = \pi \hbar^2 n/m$ in 2DES), which are however {\it
 not} completely independent of each other (since they both depend on
the electron density)--in particular, $T/T_F \sim r_s^2$ for a fixed
temperature and changing density.  

The structure of this paper goes as following: In
section~\ref{sec:theory} we present the theory for our effective mass
calculation. In section~\ref{sec:num} we provide our numerical results
of our calculated effective mass as a function of $r_s$ and $T$. In
section~\ref{sec:ana} we present the analytical results for effective
mass in the $r_s \ll 1$ and $T/T_F \ll 1$ limit. We conclude in
section~\ref{sec:con} with a brief discussion. 


\section{Theory}
\label{sec:theory}

\begin{figure}[htbp]
  \centering
  \includegraphics[width=3in]{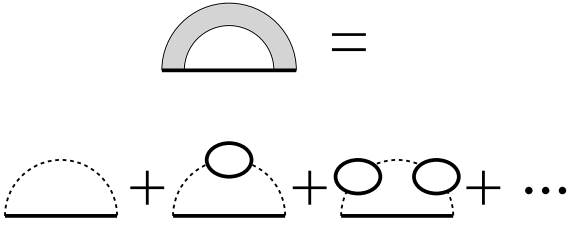}
  \caption{The Feynman diagram for the self-energy. The circles are
  polarization bubbles, the dashed lines the Coulomb interaction, and
  the solid lines the electron Green's function.}   
  \label{fig:fynman}
\end{figure}

We consider a 2DES interacting via the long range Coulomb
interaction. The effective mass renormalization is microscopically
calculated~\cite{AGD} from the electron self-energy function $\Sigma
({\bf k}, i \nu_l)$ defined at the Matsubara imaginary frequency $i
\nu_l$ and 2D wavevector ${\bf k}$. To calculate the electron
self-energy, we make the well-known ``GW''
approximation~\cite{hedin,rice,ting} of a leading order expansion in 
the dynamically screened Coulomb interaction (the corresponding
Feynman diagram for the self-energy is shown in
Fig.~\ref{fig:fynman}), obtaining ($\hbar = 1$ throughout): 
\begin{eqnarray}
  \label{eq:self}
  \Sigma({\bf k}, i \nu_l) = - \int \frac{d^2 q}{(2 \pi)^2}
    T \sum_{\omega_n} \frac{V_q}{\epsilon({\bf q}, i \omega_n)}
      \nonumber \\
      \times \frac{1}{i \nu_l + i \omega_n - \xi_{{\bf q} - {\bf k}}},
\end{eqnarray}
where $V_q = 2 \pi e^2 /q$ is the 2D bare Coulomb potential, $i
\nu_l = i (2l + 1) \pi k_B T$ and $i \omega_n = i 2 n \pi k_B T$ are
the usual fermion/boson odd/even Matsubara frequencies ($l$, $n$
integers), $\xi_{\bf k} = k^2 /(2 m) - \mu$, $\mu$ the chemical
potential, and $\epsilon({\bf k}, i \omega_n)$ is the RPA dynamical
dielectric function, given by the sum of the polarization bubble
diagrams: 
\begin{equation}
  \label{eq:epsilon}
  \epsilon({\bf k}, i \omega_n) = 1 - V_q \Pi({\bf k}, i \omega_n),
\end{equation}
with $\Pi({\bf k}, i \omega_n)$ the electronic 2D
polarizability. Within RPA, we have
\begin{equation}
  \label{eq:Pi}
  \Pi({\bf k}, i \omega_n) = 2 \int \frac{d^2 q}{(2 \pi)^2}
  \frac{n_F(\xi_{\bf q}) - n_F(\xi_{{\bf q} - {\bf k}})}
  {\xi_{\bf q} - \xi_{{\bf q} - {\bf k}} + i \omega_n},
\end{equation}
where $n_F(x) = 1/(e^{x/T} + 1)$ is the Fermi distribution
function.  

The form of retarded polarizability $\Pi({\bf k} , \omega) \equiv
\Pi({\bf k}, i \omega_n \to \omega + i 0^+)$ has been provided by
previous work~\cite{stern} at zero temperature:
\begin{eqnarray}
\label{eq:P0}
\Pi_0 ({\bf k}, \omega, \mu) 
= -{m \over \pi} + {m^2 \over \pi k^2} 
\Bigg[ \sqrt{ (\omega + {k^2 \over 2m} )^2 - {2 \mu k^2 \over 2m} } 
\nonumber \\
- \sqrt{ (\omega - {k^2 \over 2m} )^2 
- {2 \mu k^2 \over 2m} } \Bigg],
\end{eqnarray}
where $\mu$ is the chemical potential. The finite temperature form of
retarded polarizability can obtained from Eq.~(\ref{eq:P0}) by
\begin{equation}
\label{eq:PT}
\Pi ({\bf k}, \omega, \mu; T) = \int\limits_0^\infty d \mu' 
{\Pi_0({\bf k}, \omega, \mu') \over 4 T \cosh^2 ({\mu' - \mu \over 2T}) }.
\end{equation}

The quasiparticle energy $E_{\bf k}$ is obtained from the Dyson
equation using the analytically continued retarded self-energy $\Sigma
({\bf k}, i \nu_l \to \omega + i 0^+) \equiv \Sigma({\bf k}, \omega)$:  
\begin{equation}
  \label{eq:Dyson}
  E_{\bf k} = \xi_{\bf k} + \mbox{Re}\Sigma({\bf k}, E_{\bf k}).
\end{equation}
Eq.(\ref{eq:Dyson}) is exact, while our GW-approximation is the first
order perturbation expansion in the dynamically screened
interaction. There has been much discussion on whether one should use
exact Eq.~(\ref{eq:Dyson}) for the effective mass or the so-called
on-shell approximation, keeping only first order interaction terms by
taking the first order iteration of Eq.~(\ref{eq:Dyson}): 
\begin{equation}
  \label{eq:Ek}
  E_{\bf k} = \xi_{\bf k} + \mbox{Re}\Sigma({\bf k}, \xi_{\bf k}).
\end{equation}
The on-shell approximation is expected to be more accurate within the
GW scheme as it effectively accounts for some vertex corrections and 
obeys the Ward identities. This approach has previously been used in
2D~\cite{ting} and 3D~\cite{rice} zero-temperature effective mass
calculations, and is regarded to be better than solving the full Dyson equation
(Eq.~(\ref{eq:Dyson}) above). The two approaches are identical in the
high-density limit $r_s \ll 1$. For $r_s > 1$, they give qualitatively
similar but quantitatively different results. The effective mass can
then be derived from the relation  $1/m^* = \left. \left( k^{-1} d E_k
/ d k \right) \right|_{k = k_F}$, remembering that the bare band mass
$m$ is given by $1/m = \left. \left( k^{-1} d \xi_k / d k \right)
\right|_{k = k_F}$:  
\begin{equation}
  \label{eq:mass}
  \frac{m^*}{m} = \left. \left[1 + \frac{m}{k}\frac{d}{d k}
    \mbox{Re}\Sigma(k, \xi_k) \right]^{-1} \right|_{k = k_F}
\end{equation}
where $k_F$ is the Fermi momentum for the non-interacting 2DES.

We use three different techniques in calculating the self-energy:
frequency sum, frequency integration, and plasmon-pole
approximation. The first two techniques are equivalent to each other,
and correspond to different ways of doing the analytic continuation of
the imaginary frequency self-energy. The frequency sum technique is
explained in Ref.~\cite{hu}, and the frequency integration technique,
also called spectral representation, in Ref.~\cite{AGD}. In the
frequency sum method, the retarded self-energy is given by
\begin{eqnarray}
  \label{eq:Esum}
  \mbox{Re}\Sigma({\bf k}, \omega) =&-& \int \frac{d^2 q}{(2 \pi)^2}
  V_q n_F(\xi_{{\bf q} - {\bf k}}) \nonumber \\
  &-& \int \frac{d^2 q}{(2 \pi)^2} V_q \mbox{Re} \left[
    \frac{1}{\epsilon(q, \xi_{{\bf q} - {\bf k}} - \omega)} - 1 \right]
  \nonumber \\
  &&~~~~~\times \left[n_B(\xi_{{\bf q} - {\bf k}} - \omega)
    +n_F(\xi_{{\bf q} - {\bf k}}) \right] \nonumber \\
  &-& \int \frac{d^2 q}{(2 \pi)^2} T \sum_{\omega_n}
  V_q \left[ \frac{1}{\epsilon(q, i \omega_n)} - 1 \right]
  \nonumber \\
  && ~~~~ \times \frac{1}{i \omega_n - (\xi_{{\bf q} - {\bf k}} - \omega)},
\end{eqnarray}
where $n_B(x) = 1/(e^{x/T} - 1)$ is the Bose distribution
function. For the frequency integration method, the retarded
self-energy is 
\begin{eqnarray}
  \label{eq:Eint}
  \mbox{Re}\Sigma({\bf k}, \omega) =&-& \int \frac{d^2 q}{(2 \pi)^2}
  V_q n_F(\xi_{{\bf q} - {\bf k}}) \nonumber \\
  &-& \int \frac{d^2 q}{(2 \pi)^2} \int \frac{d \nu}{2 \pi} 2 V_q 
  \mbox{Im} \frac{1}{\epsilon(q, \nu) }
  \nonumber \\
  &&~~~~~\times \frac{ n_B(\nu)
    +n_F(\xi_{{\bf q} - {\bf k}}) }
  {\nu - (\xi_{{\bf q} - {\bf k}} - \omega)}.
\end{eqnarray}

The plasmon-pole approximation (PPA) is a simpler
technique~\cite{vinter,lundqvist,dassarma} for carrying out the
frequency sum in the RPA self-energy calculation by using a spectral
pole (i.e. a delta function) ansatz for the dynamical dielectric
function $\epsilon({\bf   k}, \omega)$: 
\begin{equation}
  \label{eq:PPA}
  \mbox{Im}\epsilon^{-1}({\bf k}, {\omega}) =
  C_k \left[ \delta(\omega - \bar{\omega}_k) - \delta(\omega +
  \bar{\omega}_k) \right] /2,
\end{equation}
where the spectral weight $C_k$ and the pole $\bar{\omega}_k$ of the
PPA propagator in Eq.~(\ref{eq:PPA}) are determined by using the
Kramers-Kr\"{o}nig relation (i.e. causality) and the $f$-sum rule
(i.e. current conservation). We mention that $\bar{\omega}_k$ in
Eq.~(\ref{eq:PPA}) does {\it not} correspond to the real plasmon
dispersion in the 2DES, but simulates the whole excitation spectra of
the system behaving as an effective plasmon at low momentum and as the
single-particle electron-hole excitation at large momentum, as
constrained by the Kramers-Kr\"{o}nig relation and
the $f$-sum rule. Details on the PPA are available in
literature~\cite{vinter,lundqvist}, including the finite-temperature
generalization~\cite{dassarma}. The PPA, which is
known~\cite{vinter,lundqvist,dassarma} to give results close to the
full RPA calculation of self-energy, allows a trivial carrying out of
the frequency sum in the retarded self-energy function
leading to: 
\begin{eqnarray}
\label{eq:EPPA}
  \mbox{Re}\Sigma({\bf k}, \omega) =&-& \int \frac{d^2 q}{(2 \pi)^2}
  V_q n_F(\xi_{{\bf q} - {\bf k}}) \nonumber \\
  &-& \int \frac{d^2 q}{(2 \pi)^2} V_q C_q \Big[ 
  \frac{ n_B( \bar{\omega}_q ) + n_F ( \xi_{{\bf q} - {\bf k}} ) }
  {\bar{\omega}_q - ( \xi_{{\bf q} - {\bf k}} - \omega ) } \nonumber \\
&&~~~~~~~+ 
  \frac{ n_B( -\bar{\omega}_q ) + n_F ( \xi_{{\bf q} - {\bf k}} ) }
  {\bar{\omega}_q + ( \xi_{{\bf q} - {\bf k}}  - \omega) } \Big].
\end{eqnarray}

We calculate the self-energy by carrying out the 2D momentum
integration (Eq. \ref{eq:Esum}, \ref{eq:Eint}, \ref{eq:EPPA}) as well
as the frequency sum (Eq. \ref{eq:Esum}) and the frequency integral
(Eq. \ref{eq:Eint}) in order to obtain the quasiparticle effective mass
(Eq. \ref{eq:mass}). We emphasize that our reason for carrying out our
calculation of the electron self-energy by three different techniques
(RPA frequency sum and integration, and PPA) is to completely ensure
the numerical accuracy of the calculated temperature dependent
effective mass by comparing the consistency among the three sets of
results. This is particularly significant since there is no existing
temperature-dependent effective mass calculation in the
literature for us to compare with. The fact that our three sets of
results are consistent with each other (and we reproduce the
existing~\cite{ting,vinter,jalabert,schulze,ando} $T=0$ effective mass
results from our finite temperature theory) provides compelling
support for our conclusions in this paper. Since our results obtained
in the three techniques are in good agreement, we will only show here
our effective mass results using RPA frequency sum method for the sake 
of brevity.


\section{Numerical results}
\label{sec:num}

\begin{figure}[htbp]
  \centering
  \includegraphics[width=3in]{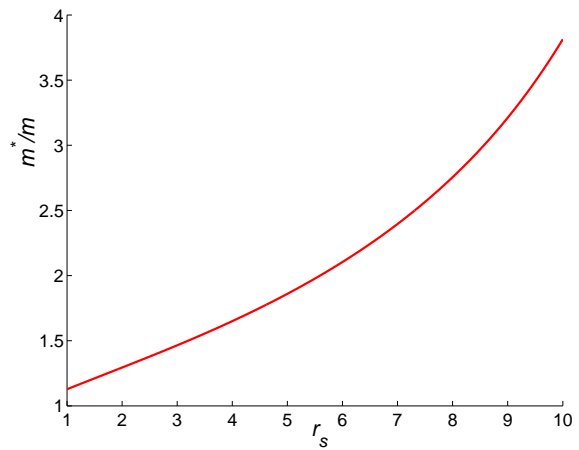}
  \caption{Calculated T=0 effective mass as a function of $r_s$ in a
  2DES in the high $r_s$ region.}
  \label{fig:zeroThrs}
\end{figure}

\begin{figure}[htbp]
  \centering
  \includegraphics[width=3in]{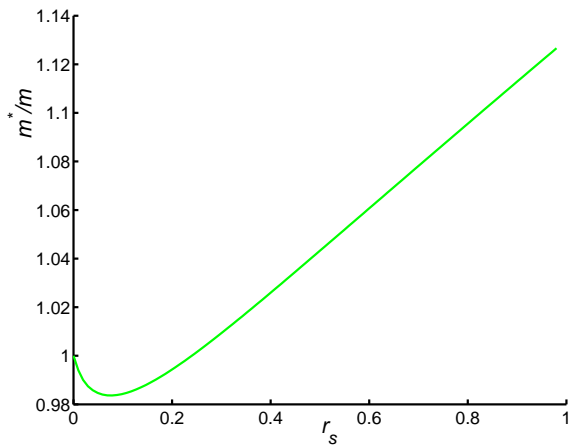}
  \caption{Calculated T=0 effective mass as a function of $r_s$ in a
  2DES in the low $r_s$ region. }
  \label{fig:zeroTlrs}
\end{figure}

First, we present our extreme low temperature 2D result ($T/T_F
\approx 10^{-4}$) in Fig.~\ref{fig:zeroThrs} and
Fig.~\ref{fig:zeroTlrs}, to be compared with the existing $T=0$ 2D
results~\cite{ting,vinter,jalabert,schulze,ando}, for $m^*(r_s)$ in
the $r_s = 0 - 10$ range, showing that the effective mass
renormalization could be almost as large as $5$ for dilute $r_s \sim
10$ 2DES. We emphasize that the results presented in
Fig.~\ref{fig:zeroThrs} and Fig.~\ref{fig:zeroTlrs} based on the $T
\to 0$ limit of our finite temperature theory are in {\it
  quantitative} agreement with the existing $T=0$ 2D RPA effective
mass calculations~\cite{ting} which were, however, restricted to the
$r_s$ ($< 5$) regime.  

\begin{figure}[htbp]
  \centering
  \includegraphics[width=3in]{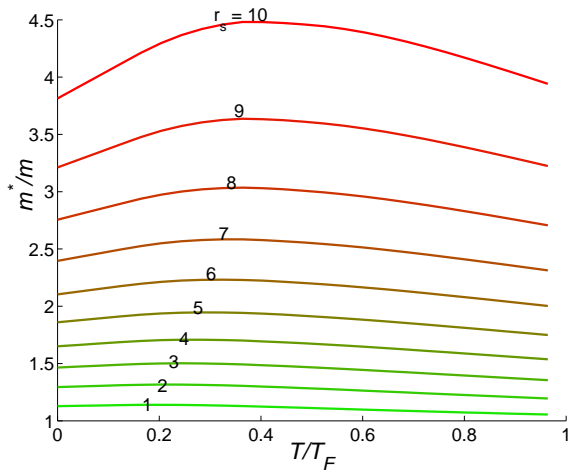}
  \caption{Calculated 2D effective mass as a function of $T/T_F$ for
  different high $r_s$ values: $r_s = 10 \to 1$ from top to
  bottom. Note that $T_F \propto r_s^{-2}$, making the absolute
  temperature scale lower for higher $r_s$ values.}
  \label{fig:mTr}
\end{figure}

\begin{figure}[htbp]
  \centering
  \includegraphics[width=3in]{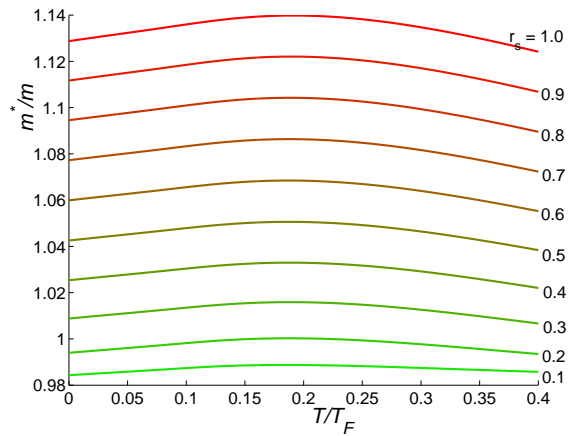}
  \caption{Calculated 2D effective mass as a function of $T/T_F$ for
  different low $r_s$ values: $r_s = 1 \to 0.1$ from top to bottom.}
  \label{fig:mTrlrs}
\end{figure}

In Fig.~\ref{fig:mTr} we show our calculated 2D $m^*(T)$ as a function
of $T/T_F$ for different values of the 2D interaction parameter $r_s$
($=1-10$). Fig.~\ref{fig:mTrlrs} shows the effective mass temperature
dependence at high densities. In the low temperature region the
effective mass first rises to some maximum, and then decreases as
temperature increases. This peculiar behavior is present at all
densities. The nonmonotonic trend is systematic, and the value of
$T/T_F$ where the effective mass peaks increases with increasing
$r_s$. The initial increase of $m^*(T)$ is linear in $T/T_F$ as $T \to
0$, and the slope $\frac{d(m^*/m)}{d(T/T_F)}$ is almost independent of
$r_s$ for very small $r_s$ ($<1$), but increases with $r_s$ for larger
$r_s$ values. We mention that we get somewhat stronger temperature
dependence (i.e. larger $dm^*/dT$) in our PPA calculation (not shown
here).

\begin{figure}[htbp]
  \centering
  \includegraphics[width=3in]{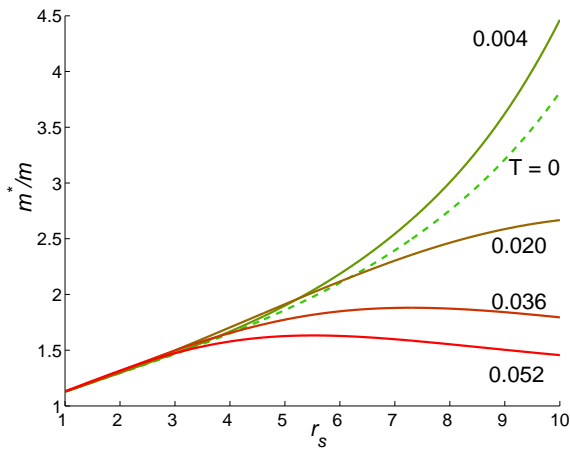}
  \caption{Calculated $m^*/m$ at fixed values of temperatures. $T$ is in
  the units of $T_F$ at $r_s = 1$.}
  \label{fig:mrT}
\end{figure}

In Fig.~\ref{fig:mrT} we show the dependence of the effective mass
renormalization as a function of the interaction parameter $r_s$ for a
few values of {\it fixed} temperature (rather than fixed $T/T_F$,
remembering that $T_F \propto r_s^{-2}$ since $T_F \propto n$ and $r_s
\propto n^{-1/2}$). The calculated $m^*(r_s)$ for fixed $T$ values is
quite striking: For low fixed values of $T$, $m^*/m$ initially
increases with $r_s$ even faster than the corresponding $T=0$ result,
eventually decreasing with $r_s$ at large enough values (where the
corresponding $T/T_F$ values become large enough). This nonmonotonic
behavior of $m^*(r_s)$ as a function of $r_s$ for fixed temperatures
showing a temperature-dependent maximum (with the value of $r_s$ at
which the $m^*$ peak occurs decreasing with increasing $T$ as in
Fig.~\ref{fig:mrT}) is complementary to the nonmonotonicity of
$m^*(T)$ in Fig.~\ref{fig:mTr} as a function of $T/T_F$ (at fixed
$r_s$) and arises from the relationship between the dimensionless
variables $T/T_F$ ($\propto r_s^{-2}$) and $r_s$ ($\propto
T_F^{-1/2}$) due to their dependence on the carrier density (i.e. $T_F
\propto n$ and $r_s \propto n^{-1/2}$). 

One immediate consequence of our results shown in Figs.~\ref{fig:mTr}
and \ref{fig:mrT} is that $m^* (T/T_F, r_s) \equiv m^*(T, n)$ in 2DES
could show a strong enhancement at low (but finite) temperatures and
low electron densities (large $r_s$). Comparing with the actual system
parameters for 2D electrons in Si inversion
layers~\cite{shashkin1,shashkin2} and GaAs heterostructures~\cite{zhu}
(and taking into account the quasi-2D form factor effects~\cite{ando}
neglected in our strictly 2D calculation) we find that, consistent
with recent experimental findings~\cite{shashkin1,shashkin2,zhu}, our
theoretical calculations predict (according to Figs.~\ref{fig:mTr} and
\ref{fig:mrT} as modified by subband form factors) $m^*/m$ to be
enhanced by a factor of $2-4$ for the experimental densities and
temperatures used in recent
measurements~\cite{shashkin1,shashkin2,zhu}. Due to the approximate
nature of our theory we do not further pursue the comparison with
experimental data in this paper leaving that for a future study.  

\section{Analytical results in $r_s \ll 1, T/T_F \ll 1$ limit}
\label{sec:ana}

We have also carried out an analytic calculation of the
temperature-dependent 2D effective mass in the leading order
dynamically screened interaction. This turns out to be an extremely
difficult task due to the highly complicated nonanalytic structure of
the integrand in Eq.~(\ref{eq:self}), or equivalently
Eqs.~(\ref{eq:Esum}) or (\ref{eq:Eint}). It is only possible to carry
out our analytical work in the high-density ($r_s \ll 1$), low
temperature ($T/T_F \ll 1$) limit. It is well established that at zero
temperature one can do an $r_s$ expansion for the quasiparticle
self-energy in the $r_s \ll 1$ limit since the leading order
contribution in $r_s$ comes only from the ring diagrams, which are
exactly what we calculated in our theory (Fig.~\ref{fig:fynman}). In
this sense, RPA is {\em exact} in the high-density limit, and one can
calculate an exact quasiparticle effective mass from the diagrams of
Fig.~\ref{fig:fynman} in the $r_s \to 0$ limit. In the current work we
carry out an exact expansion in the $r_s \to 0$ and $T/T_F \to 0$
limit. At finite temperatures, our analysis shows that it is valid to
expand self-energy and effective mass in $r_s$ and $T/T_F$ in the $r_s
\ll 1, T/T_F \ll 1$ region. Again we prove that the leading order
contribution in $r_s$ and $T/T_F$ also only comes from the ring
diagrams, and therefore RPA remains {\em exact} in this limit. Out
calculation shows that 
\begin{equation}
  \label{eq:analytic}
  \frac{m^*}{m} = A(r_s) + B(r_s) \left( {T  \over T_F} \right)
  + C (r_s) \left( \frac{T}{T_F} \right)^2 \ln \left( \frac{T_F}{T} \right),
\end{equation}
where $A(r_s)$, $B(r_s)$ and $C(r_s)$ are functions independent of
termpature, and $B(r_s) \approx B_0 > 0$ and $C(r_s) < 0$ for $r_s \ll
1$.  

In Eq.~(\ref{eq:analytic}) the first term $A(r_s)$ is responsible for
the nonmonotonic behavior of $r_s$ dependence of $m^*$ at zero
temperature, i.e. when $T=0, r_s \ll 1$, effective mass first deceases
with increasing $r_s$, and then increases. This corresponds to our
zero temperature effective mass curve as a function of $r_s$ in $r_s
\ll 1$ region, which is shown in Fig.~\ref{fig:zeroTlrs}. We mention
here that this nonmonotonic zero temperature $r_s$ dependence of the
effective mass has already been found by previous
works~\cite{ting,vinter,jalabert,schulze,ando}.    

The second term in Eq.~(\ref{eq:analytic}) accounts for the
leading-order temperature correction to the effective mass, which is
of more interest to us. Our analytical calculation shows that $B(r_s)
\approx B_0 > 0$ for $r_s \ll 1$, ensuring that the leading-order 
temperature correction, in agreement with our numerical 
results, enhances the effective mass renormalization in a linear
manner as $T \to 0$. Moreover, the linear temperature coefficient is
independent of $r_s$ as $r_s \ll 1$ is also in good agreement with our
numerical results in this region, as shown in Fig.~\ref{fig:mTrlrs}
and discussed in section~\ref{sec:num}. 

The subleading temperature correction, shown as the third term in
Eq.~(\ref{eq:analytic}), is negative. This correction, combining with the
leading order linear temperature correction, produces the peak in
the effective mass temperature dependence. This again agrees
qualitatively very well with our numerical findings in the $r_s \ll 1,
T/T_F \ll 1$ region as shown in Fig.~\ref{fig:mTrlrs}. Of course it is
very difficult to determine whether the subleading temperature
correction is $T^2$ or $T^2 \ln T$ dependence just by examining the
numerical results, but the sign of this correction is certainly
negative.  


\section{Conclusion}
\label{sec:con}

Our most important new result is the unanticipated non-analytic
linear-$T$ enhancement of the quasiparticle effective mass at low
$T/T_F$ and for {\it all} densities. This result transcends our
specific GW approximation scheme since it persists for $r_s \ll 1$
where our approximation is exact. Since all quantum many-body
renormalization must vanish in the classical high temperature limit,
it follows rigorously that $m^*(T)$ must be nonmonotonic with a peak
somewhere at an intermediate temperature as shown in
Fig.~\ref{fig:mTr}. We point out, however, that this nonmonotonicity
would not be easy to observe experimentally since the quasiparticle is
unlikely to be well-defined at finite values of $T/T_F$ ($\sim 0.2 -
0.8$) where the peak of $m^*(T)$ lies. On the other hand, it should be
possible to experimentally verify our predicted non-analytic linear in
$T$ enhancement of the quasiparticle effective mass at low $T/T_F$.

Finally, we comment on the approximations used in our
calculation. First, our theory leaves out quasi-2D form factor (and
related solid state physics) effects which are straightforward to
include~\cite{ando} by appropriately modifying the bare interaction
$V_q$ in the theory, and would not lead to any qualitative changes in
the results (but would reduce the magnitude of the mass
renormalization by a factor of $1.2$ to $2$ depending on the electron
density). Second (and more importantly), our use of the leading-order
GW-RPA approximation, which is exact only in the high density ($r_s
\ll 1$) limit, is open to question. Although we believe that at finite
temperatures the GW-RPA approximation becomes more accurate (and our
quasiparticle energy calculation of Eq.~(\ref{eq:Pi}) approximately
incorporates some vertex corrections going beyond the leading order
expansion in the dynamically screened interaction~\cite{rice,ting}),
our principal rationale for carrying out the GW-RPA many-body
calculation is that (a) it is the {\it only systematic} many-body
perturbative calculation that is feasible for interacting quantum
Coulomb systems; and (b) RPA, while being exact only in the weakly
interacting $r_s \ll 1$ limit, is known to produce qualitatively
reasonable results even in the strongly interacting ($r_s > 1$)
regime, as demonstrated by the agreement between RPA and experiments
in 3D metals ($r_s \approx 3 - 5$) and in 2D semiconductor systems
($r_s \approx 1 - 10$). The fact that  our predicted nonanalytic low
temperature many-body enhancement of effective mass systematically
persists to the $r_s \ll 1$ regime shows the generic validity of our
results. In additiion, RPA self-energy calculation should become more
accurate as $r_s$ increases (i.e. decreasing density) for a fixed
non-zero $T$ because RPA is exact at any density for $T/T_F \gg 1$.

In this context we emphasize that the RPA self-energy calculation
(i.e. our effective mass calculation based on the diagrams of
Fig.~\ref{fig:fynman}) is an expansion in the dynamically screened
Coulomb interaction which becomes equivalent to an expansion in $r_s$
only in the $r_s \to 0$ limit. The RPA self-energy at arbitrary $r_s$
may {\em not} be an expansion in $r_s$ at all, but in some other
effective parameters. Even in the high-density $r_s \to 0$ limit, the
effective expansion parameter turn out to be $r_s /\gamma$ where
$\gamma$ is a number of order 15(5) in 3(2) dimensional systems.

This work is supported by NSF-ECS, ONR, DARPA, and LPS.


\end{document}